\documentclass[iop]{emulateapj}

\usepackage{natbib}
\usepackage{graphicx}
\usepackage{amsmath}
\usepackage{hyperref}
\usepackage{afterpage}

\newcommand{\vh}{v_{\rm helio}}
\newcommand{\vsys}{v_{\rm sys}}

\newcommand{\mvir}{M_{\rm vir}}

\newcommand{\feh}{[{\rm Fe}/{\rm H}]}

\newcommand{\Lsun}{L_\odot}
\newcommand{\Msun}{M_\odot}

\newcommand{\rhalf}{r_{\rm 1/2}}
\newcommand{\slos}{\sigma_{\rm LOS}}
\newcommand{\kps}{km~s$^{-1}$}

\newcommand{\ate}{And~XXVIII}
\newcommand{\atn}{And~XXIX}
\newcommand{\kinfigs}{Figures \ref{fig:and28kin} and \ref{fig:and29kin}}
\newcommand{\citeslbellt}{\citet{Slater11And28} and \citet{Bell11And29}}
\newcommand{\besancon}{Besan\c{c}on}
\newcommand{\atevwindow}{$-350$ to $-300$ \kps}
\newcommand{\atnvwindow}{$-215$ to $-175$ \kps}

\newcommand{\Ss}{\S }

\begin{document}

\title{The Outer Limits of the M31 System: Kinematics of the Dwarf Galaxy Satellites And XXVIII \& And XXIX\footnotemark[*]} 
\shorttitle{Outer Limits of M31}


\author{Erik J. Tollerud\altaffilmark{1,2,3}, Marla C. Geha\altaffilmark{1}, Luis C. Vargas\altaffilmark{1}, James S. Bullock\altaffilmark{2} 
}
\altaffiltext{1}{Astronomy Department, Yale University, P.O. Box 208101, New Haven, CT 06510, USA; erik.tollerud@yale.edu, marla.geha@yale.edu, 
luis.vargas@yale.edu}
\altaffiltext{2}{Department of Physics and Astronomy, 4129 Frederick Reines Hall, University of California, Irvine, Irvine, CA, 92697, USA; bullock@uci.edu} 
\altaffiltext{3}{Hubble Fellow}

\footnotetext[*]{The data presented herein were obtained at the W.M. Keck Observatory, which is operated as a scientific partnership among the 
California Institute of Technology, the University of California and the National Aeronautics and Space Administration. The Observatory was 
made possible by the generous financial support of the W.M. Keck Foundation}

\begin{abstract}

We present Keck/DEIMOS spectroscopy of resolved stars in the M31 satellites
 \ate{} \& \atn{}.   We show that these are likely self-bound galaxies based
on 18 and 24 members in \ate{} \& \atn{}, respectively.   \ate{} has
a  systemic velocity of $-331.1 \pm 1.8$
\kps{} and velocity dispersion of  $4.9\pm1.6$  \kps{}, implying a
mass-to-light ratio (within $\rhalf$) of $\sim 44 \pm 41$.   \atn{} has a systemic velocity
of $-194.4 \pm 1.5$ \kps{} and velocity dispersion of $5.7\pm1.2$ \kps,
implying a mass-to-light ratio  (within $\rhalf$)  of $\sim 124 \pm 72$.    
The internal kinematics and implied masses of \ate{} \& \atn{} are similar to
dwarf spheroidals (dSphs) of comparable luminosities, implying that these objects 
are dark matter-dominated dwarf galaxies. Despite the large projected distances from 
their host ($380$ and $188$ kpc), the kinematics
of these dSph suggest that they are bound M31 satellites.

\end{abstract}

\section{Introduction}
\label{sec:intro}

Local Group (LG) dwarf galaxies are crucial elements of near-field cosmology.  They serve as valuable probes of low luminosity galaxy formation, and 
provide an important window into the nature of satellite-host interactions  \citep[e.g.,][]{bullock00, stri07, krav10satrev, kaz11stirr, BKBK11, 
anderhalden12}. Their utility stems primarily from their proximity and low surface densities, allowing spectroscopic observations of  individual 
stars.  This enables far more in-depth studies than are possible for more distant targets. 

The numbers of known M31 and Milky Way (MW) dSph satellites has increased greatly in recent years.  This is  predominantly due to 
homogenous wide-field surveys, particularly  the Sloan Digital Sky Survey (SDSS) for the MW \citep{willman05, bel07}, and  the Pan-Andromeda Archeological Survey   \citep[PAndAS,][]{ibata07, pandas09nat}.  
The environs of M31 have been particularly fertile ground for discovering dSphs, yielding 20 of its 28 known dSphs in the last six years. 

Two of the most  recently discovered M31 satellite candidates are \ate{}, found by \citet{Slater11And28}, \& \atn{},  by 
\citet{Bell11And29}. 
Projected in the plane of the  sky at the distance of M31, \ate{} \& \atn{} lie at $380$ and $188$ kpc, respectively. 
The 3D distances from M31 are $367$ and $188$ kpc, using the tip of the red giant branch (TRGB) line-of-sight 
distances of \citeslbellt{}.  The expected virial radius for M31's dark matter halo is $\sim 
300$ kpc \citep{klypin02, watkins10}, so \atn{} lies in the outskirts of M31's halo and \ate{} just outside.  
This makes \ate{} the second-most distant M31 dSph\footnote{The starforming dIrrs IC1613 and Peg are 
more distant and possibly associated with M31\citep{mcc12LGcat}.} satellite, 
with only And XVIII more distant (see \S \ref{ssec:satorfield}).   

\ate{}, in particular, is potentially an important data point for galaxy formation.  It may be an analog of Leo T, a satellite in the outskirts of the MW halo and the 
only low-luminosity MW satellite known to contain HI gas \citep{irwin07leot, ryan-weber08}.  
This motivates more detailed studies of these M31 satellites, as Leo T has been valuable
for understanding star formation and gas  stripping in dSphs \citep[e.g.,][]{ricotti09, GP09, bovill11, nichols11}. If they have no recent star 
formation, \ate{} \& \atn{} may be more akin to Tucana and Cetus.  These are unusual, non-starforming dSphs of the LG that 
do not seem to be satellites \citep{lavery92, whiting99, monelli10}, but may be ``backsplash'' galaxies that were once much closer to either M31 or the MW 
\citep{knebe11,oman13}.

Here we present the first spectroscopic observations of \ate{} \& \atn{},  obtaining kinematics of resolved stars using the DEIMOS 
spectrograph on the Keck II Telescope.
  In \S \ref{sec:obs}, we describe our observations and analysis procedures,
 in \S \ref{sec:res} we present our kinematical 
measurements for \ate{} \& \atn{}, and in \S \ref{sec:conc} we
summarize and conclude.  
Throughout this paper we adopt a distance  modulus to M31 of $\mu_{\rm M31}=24.47$, corresponding to a distance of $783$ kpc 
\citep{Mcconnachie05}, and distances of $650^{+150}_{-80}$  ($\mu=24.06 ^{+0.5}_{-0.2}$) and $730 \pm 75$ kpc ($\mu=24.31 \pm 0.22$) for \ate{} \& \atn{}, respectively \citep{Slater11And28, Bell11And29}.

\section{Observations and Analysis}
\label{sec:obs}

\begin{deluxetable*}{lllll}[!tbh]
\tabletypesize{\tiny}
\tablecaption{Photometric and Kinematic Properties of And 28 \& And 29}
\tablehead{
\colhead{Row} & \colhead{Quantity} & \colhead{Units} &
\colhead{\ate} & \colhead{\atn} \\ 
}
\startdata
(1) & $\alpha$ (J2000)      &  h$\,$:$\,$m$\,$:$\,$s  &  $22^{\rm h}32^{\rm m}41\fs 2$   & $23^{\rm h}58^{\rm m}55 \fs 6$ \\
(2) & $\delta$ (J2000)       &  $^\circ\,$:$\,'\,$:$\,''$   & $31\arcdeg 12 \arcmin 58.2 \arcsec$  & $30 \arcdeg 45 \arcmin 20.0 \arcsec$  \\
(3) & $d_{\rm LOS}$   & kpc  & $650^{+150}_{-80} $              & $730 \pm 75$ \\
(4) & $M_{V}$        & mag  & $-8.5 ^{+0.4}_{-1.0}$      & $-8.3 \pm 0.4$   \\   
(5) & $R_{\rm eff}$  &   arcmin   &  $1.11 \pm 0.21$   &  $1.7 \pm 0.2$ \\
(6) & $R_{\rm eff}$  &   pc          &  $210^{+60}_{-50}$   &  $360 \pm 60$ \\
(7) & \feh          &  dex  &  $\sim -2.0$ & $\sim -1.8$ \\
(8) & $r_{1/2}$  &   pc          &  $280^{+80}_{-67}$   &  $480 \pm 80$ \\
\hline
\\[-0.5em]
\multicolumn{5}{c}{Keck/DEIMOS Results}\\
(9)   & $N_{\rm member}$        &             & 18    & 24 \\
(10) & $v_{\rm sys}$               &  \kps  &   $-331.1 \pm 1.8$ & $-194.4 \pm 1.5$\\
(11) & $\sigma_{\rm LOS}$    & \kps &      $4.9\pm1.6$      &$5.7\pm1.2$  \\
(12) & $M/L_V (<\rhalf)$  &    $\Msun / \Lsun$      & $44 \pm 41$ & $124 \pm 72 $ \\
(13) & $M_{\rm 1/2}$  &    $\Msun  \times 10^6$      & $4.7 \pm 3.2$ & $11.1 \pm 4.9$
\enddata
\tablecomments{\tiny (1-7) Right ascension, declination, line-of-sight distance, absolute
  magnitude, effective radii, and metallicity are taken from \citet{Slater11And28}
  and \citet{Bell11And29} for And 28 \& And 29, respectively.  }
\label{tab:props}
\end{deluxetable*}

\subsection{Target Selection and Reduction}

We selected stars for spectroscopic follow-up from the SDSS Data Release 8 photometric
catalogs \citep{sdssdr8}. We selected stars near \ate{} \& \atn{}, eliminating objects where
the SDSS {\it model}  and {\it psf} magnitudes were
discrepant by more than $0.25$ mags.  
We then selected candidate dSph
stars near the red giant branch (RGB) of a fiducial
isochrone in the $g-i$, $r$ color-magnitude diagram (CMD).  We used 12
Gyr Dartmouth isochrones \citep{dartisoc} offset by  \citet{sfd98dust} extinctions 
with \feh{} and distance modulus matched
 to \ate{} \& \atn{} \citep{Slater11And28,Bell11And29}. 
We then populated the slitmask by selecting stars in the following 
sequence, prioritizing brighter stars for each group: those within 0.3 
mags of the isochrone, those within 0.6 mags, 
and all remaining stars. The resulting selections covered a wide enough
area in the CMD that our results do not depend strongly on the assumed
distances and \feh{} from \citeslbellt{}.

\subsection{Observations and Reduction}

Spectroscopic observations were obtained on the nights of April 22--23
and September 16--17, 2012 using the DEIMOS spectrograph on the Keck
II telescope \citep{faber02}.  We used the $1200$ lines mm$^{-1}$
grating covering a wavelength region of $6400 - 9100$ \AA{}.  This provided
a FWHM resolution of $\approx 1.3$ \AA{}, equivalent to 50 \kps{}
at the center of our wavelength range. We observed 3 slitmasks for 
\ate{}, and 2 slitmasks for \atn{}, with an average exposure time of 3200 sec per mask.

\begin{figure}[htbp!]
\epsscale{1.2}
\plotone{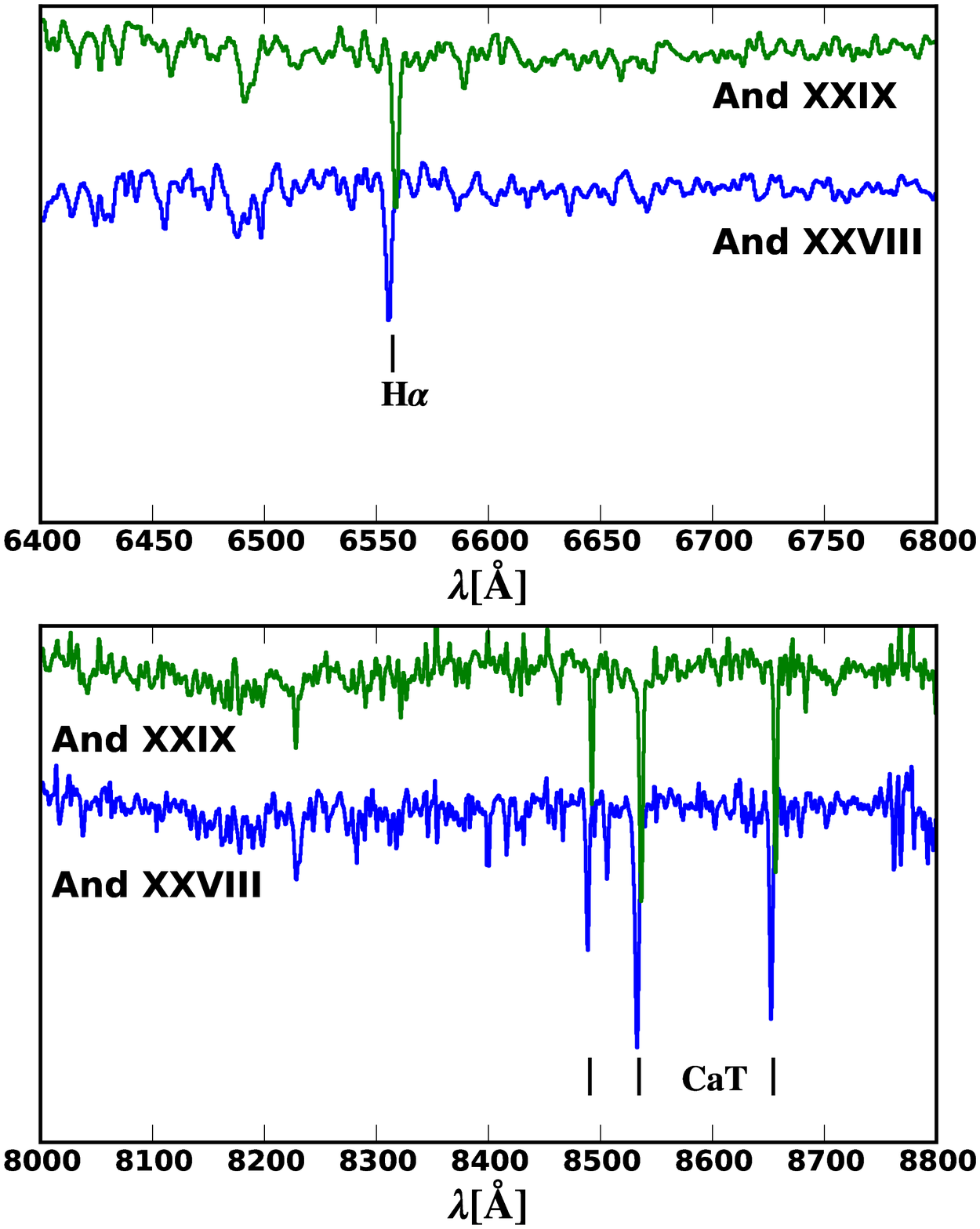}
\caption{Continuum-normalized spectra of \ate{} \& \atn{} member stars.  
These spectra are inverse variance weighted  heliocentric frame co-adds of 18 and 24 stars for \ate{} \& \atn, respectively, and 
have been smoothed with a 3-pixel gaussian filter for clarity.
The lower (blue) line is for \ate{}, while the upper (green) line is \atn{}.  The top panel is
a zoom near H$\alpha$, while the lower panel is a window that includes the calcium triplet (CaT).
  See \Ss \ref{ssec:memb} and \ref{sec:res} for details of membership determination. 
  The vertical black lines marking H$\alpha$ and the CaT are at wavelengths where the features would
  be for a velocity halfway between the $\vsys$ of \ate{} \& \atn{}.
   The offsets of the features in the spectra from these wavelengths are due to the differing $\vsys$ of the dSphs.}
\epsscale{1}
\label{fig:specs}
\end{figure}

Our spectroscopic reductions closely follow the procedure outlined in
\citet{paper1}.  We reduce the 
spectra using the spec2d pipeline developed for the DEEP2 survey, which 
produces 1d spectra from the raw images \citep{davis03deep2, spec2dascl, newman12deep2}. 
In Figure \ref{fig:specs}, we show smoothed co-adds of the resulting spectra for members stars of \ate{} 
\& \atn{}  (see \Ss \ref{ssec:memb} and \ref{sec:res} for  membership determination). The average 
signal-to-noise ratio for these spectra was 4.8 per pixel.

We cross-correlate these 1d spectra with a series of high signal-to-noise ratio (SNR) templates of known radial 
velocity to determine a line-of-sight velocity. We determine errors on these velocities by re-simulating each 
spectrum 1000 times with noise added to  simulate the per-pixel variance.  
We measure a cross-correlation velocity for each re-simulation, and use the mean 
and standard deviation of the  distribution of velocities as the line-of-sight velocity 
and uncertainty for each star \citep{sandg}. 
We also add in quadrature a systematic floor to the uncertainty, 
determined by repeat measurements of stars from  \citet{sandg}, \citet{kalirai10}, and \citet{paper1}.  
These each found consistent values of  of $2.2$ \kps{} for the systematic floor over a temporal baseline longer
than the time from these observations to \citet{paper1}.  Hence, we apply this same floor to the measurements
in this work.

\subsection{Membership}
\label{ssec:memb}

A clean sample of member stars is crucial for determining the internal
kinematics of \ate{} \& \atn{}.  The two main sources of contamination for 
these observations are foreground MW stars and M31 halo stars.  As 
demonstrated below, the MW foreground star velocity distribution 
is disjoint from the dSphs in the direction of \ate{} \& \atn{}, and hence the 
MW contaminants can mostly be  filtered with velocity cuts.  
\ate{} \& \atn{} are far from M31 in projection  ($\sim 380$ and $\sim 190$ 
kpc projected, respectively), and thus the M31 halo dominates  over the 
bulge or disk component \citep{courteau11, gilbert12}. We demonstrate in \S \ref{ssec:kin} 
that even the M31 halo density is low enough  that its contamination is unlikely to
affect our kinematic results for \ate{} \& \atn{}.  

For both \ate{} \& \atn{}, our membership determination begins by
selecting stars in the $r-i$, $r$ CMD. We select 
stars in a CMD box that encompasses stars near PARSEC isochrones
\citep{bressan12} for a 12 Gyr population with \feh{} values that match the estimates from \citeslbellt{}. 
These boxes are shown in the upper-left panels  of \kinfigs{}. We place the upper edge of these boxes above the tip of 
the isochrone because this accepts TRGB stars even if the \ate{} \& \atn{} distance moduli  are $\sim 1 \sigma$ off.
To filter out the  MW foreground stars that  lie near the dSph members in the CMD, we impose a velocity window centered 
on the ``cold spikes'' apparent in  the lower-left panels of \kinfigs. The stars within these peaks and inside the 
CMD box are spatially concentrated near the centers of \ate{} \& \atn{}, as expected for a self-bound galaxy.  Additionally,
the absence of stars near these velocities above the TRGB of our chosen isochrones show even $2\sigma$ errors in the distances 
to \ate{} \& \atn{} would have no effect on our membership determination. 

The  velocity distributions of these candidate members are consistent with a 
Gaussian in the sense that both the  \citet{shapirotest} and \citet{ADtest} 
tests cannot reject the null hypothesis of Gaussianity at the $p=0.05$ level.
While these Gaussian peaks imply that most of the selected stars are members, we 
cannot discount the possibility of a small number of contaminants that by chance are 
near the dSph locus in both the CMD and velocity.  
To address this, we estimate the surface density 
of MW foreground stars expected in our spectroscopic sample using the \besancon{} model 
of the MW \citep{besancon}.   We select model stars in the direction of each dSph which fall 
within the range of our observations in the $r-i$, $r$ CMD.  We then normalize the resulting 
model population such that the overall number of stars in the  model match our observations 
for $v_{\rm helio} > -100$ \kps, far from the dSph velocity.  
Finally, we determine how  many stars in the model would fall within the velocity window for 
the dSph members.  For both  \ate{} \& \atn{} this number is small, $\sim0.7$  and $\sim 0.8$, 
respectively, in the entire field of our observations.

\begin{figure*}[htbp!]
\plotone{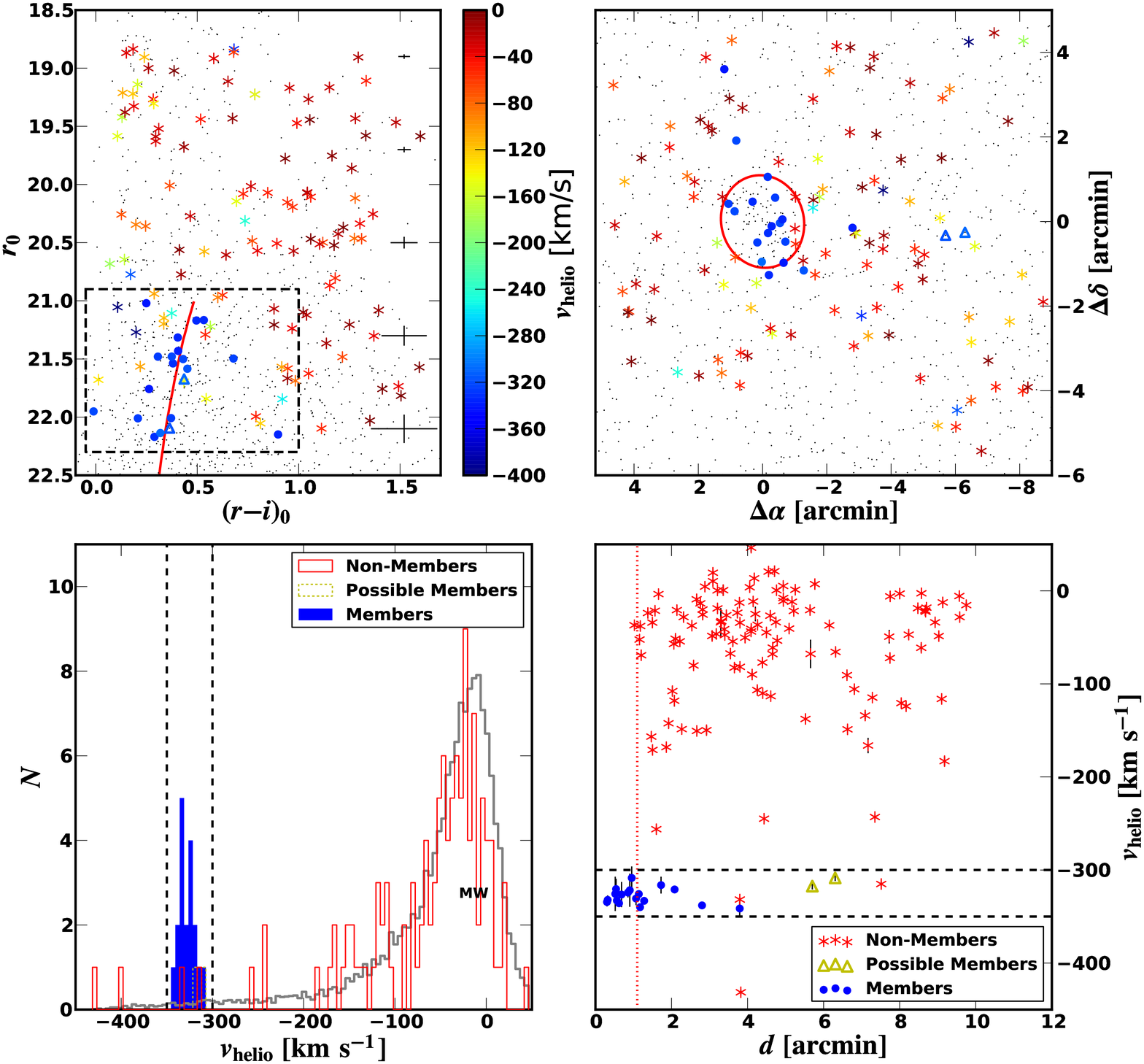}
\caption{
{\bf Upper-left:} Foreground extinction-corrected $r$, $r-i$ CMD of the SDSS photometry of stars in the \ate{} field. 
The points (black) are all star-like objects in the SDSS catalog in the spatial and CMD regions shown.  Stars with spectroscopic data 
are colored by $\vh$, with solid circles representing stars that are classified as members, 
outlined triangles are possible members (see text in \S \ref{ssec:ate}), and 
asterisks are non-members.  The (black) dashed box indicates the CMD selection  window for membership, and the black error bars at $r-i=1.5$ 
are photometric errors in the SDSS averaged in magnitude bins.  The solid (red)  line is a PARSEC isochrone of 12 Gyr age 
and \feh$=-1.5$ \citep{bressan12}.  
{\bf Upper-right:} Spatial Distribution of stars in the \ate{} field.  Symbols here are the same as those in the upper-left panel. The solid (red) line 
indicates the half-light ellipse. 
{\bf Lower-left:} Velocity histogram for stars in the \ate{} field.  The blue histogram is  member stars, while red is for 
non-members.  The (black) dashed vertical lines indicate the velocity window used for membership.  The  gray  line is the histogram 
of the  \citet{besancon} model of MW foregrounds stars, normalized as described in \S \ref{ssec:memb}.
{\bf Lower-right:} Radial velocity vs. distance from dSph in the \ate{} field. Filled circles (blue) are members, outlined triangles (yellow) are the two 
stars of uncertain status, and star-shaped symbols (red) are non-members.  Solid vertical lines are per-star velocity uncertainties.  The (black) 
dashed line is the velocity window used for determining  membership, while the (red) dotted line indicates the half-light radius of the dSph. 
}
\label{fig:and28kin}
\end{figure*}

\begin{figure*}[htbp!]
\epsscale{1}
\plotone{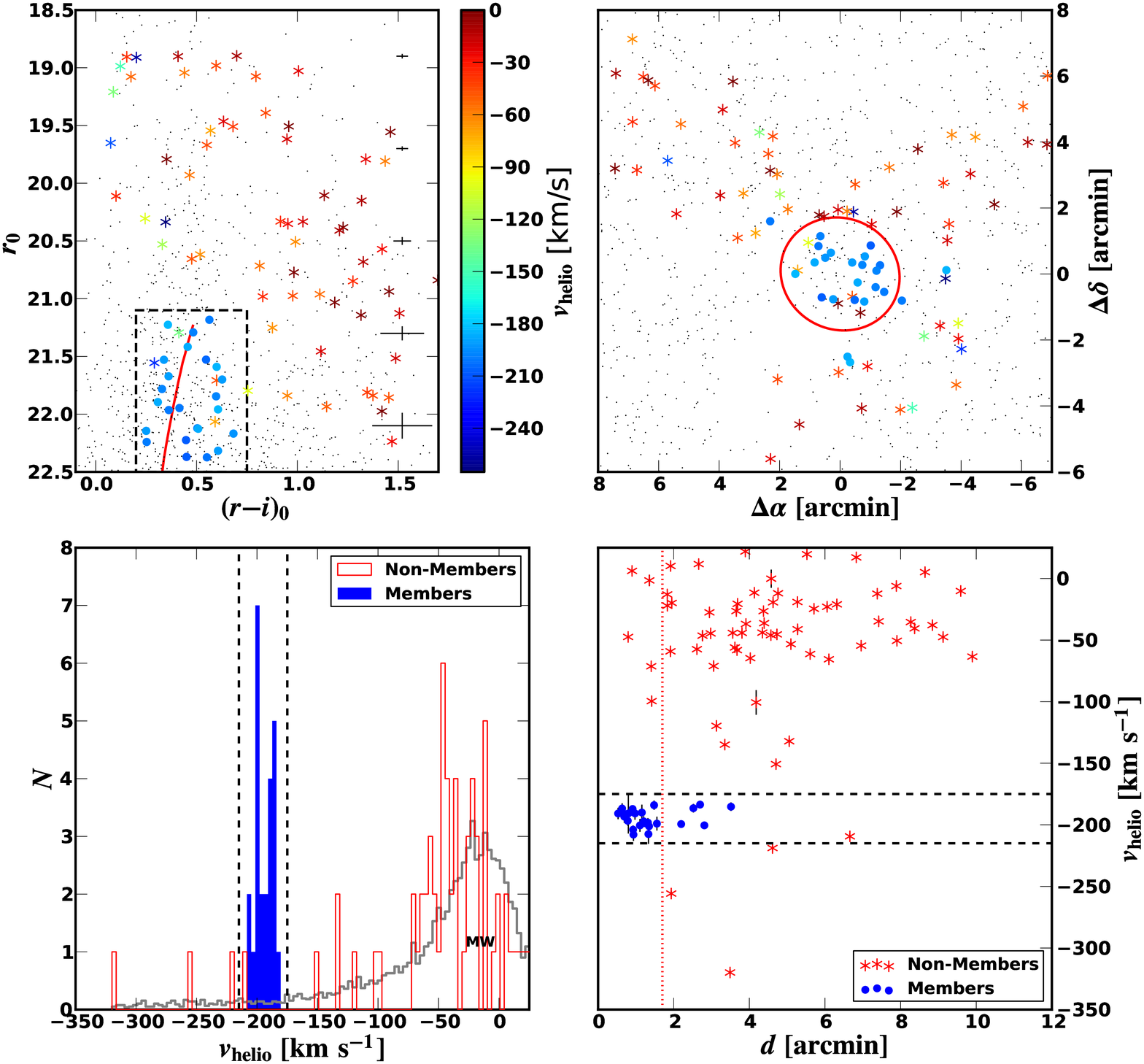}
\caption{Same as Figure \ref{fig:and28kin}, but for \atn{}.  }
\label{fig:and29kin}
\end{figure*}

\subsection{Kinematical Parameters}
\label{ssec:kin}

With a member sample selected as described above, we proceed to
determine the internal kinematics of \ate{} \& \atn{}.  In the sections
below, we model the velocity distribution of the member stars as 
a Gaussian with mean $\vsys$ (systemic velocity), and dispersion $\slos$.  
We determine the kinematical parameters of this model with a 
maximum likelihood fit, weighting all members equally,
 and use the inverse of the Hessian matrix to estimate the uncertainties on these parameters \citep{walker07, paper1}.
We examine the resulting likelihood maps for $\vsys$ and $\slos$, and note that they are very close to Gaussian
within $1\sigma$ of the peak.  This validates the use of the Hessian
matrix for estimating uncertainties, as it implicitly assumes the 
likelihood function near the maximum likelihood peak 
is approximately normal.

This approach assumes each star is only a tracer of the internal kinematics of the galaxy, and hence does not
account for the impact of binary stars.  As detailed in \citet{minor10bin} and \citet{mcconn10bin}, unresolved binaries 
can inflate the internal velocity dispersions measured for dSphs.  Repeat velocity measurements 
are required to provide any meaningful correction for this effect, and this is not available for our data set.  
However, the uncertainties we report for \ate{} \& \atn{} below are larger than the corrections that result
from multi-epoch observations of MW dSphs \citep{minor10bin, martinez11, minor13}.  Hence, unless M31
dSphs have substantially different binary populations than MW dSphs, is it unlikely unresolved binaries would have a major
effect on the results we present here.

We also estimate the M31 halo surface density near \ate{} \& \atn{}. 
 We start by determining stellar luminosity functions  from PARSEC isochrones for \feh{} $=-0.5$ and $-1.5$,
a range that approximately samples the metallicity range of the M31 halo \citep{gilbert12}.  We use these luminosity functions to 
determine how many stars bright enough to meet our spectroscopic survey limits ($r \lesssim 22.5$) are present for a
given surface brightness.  For each dSph we then determine the M31 halo surface brightness using the best-fit 
power law from  \citet{gilbert12}.  Combining these surface brightness estimates with the luminosity function thus
provides an estimate of the number of M31 halo stars expected in our fields.
We estimate $\sim 0.3$ M31 halo stars in our entire \ate{} field, and $\sim 2$ M31 halo stars in our \atn{} observations. 
Furthermore, the M31 halo velocity distribution, 
while overlapping with the dSphs, is much hotter \citep[$\sigma \sim 100$ \kps,][]{chapman06},  further reducing the 
likelihood of an M31 halo star lying within our velocity window.  
While the predicted M31 halo and MW foreground contamination is small, we nonetheless simulate its effects in \S \ref{ssec:contamination}, 
showing that it should have no impact on our inferred kinematical results.

\vfill

\section{Kinematic Results}
\label{sec:res}

\subsection{\ate{}}
\label{ssec:ate}

We present our observations of stars in the \ate{} field in Figure \ref{fig:and28kin}. This object is far from M31 ($\sim 380$ kpc projected), but is 
relatively close to the Galactic plane ($b\sim-23 \deg$).  There is a clear velocity peak at $\vsys \sim -300$ \kps{} (lower-right panel of Figure 
\ref{fig:and28kin}), well away from the velocity peak for  MW stars
($\vh \sim 0$). Applying our membership and kinematical analysis
described in \S \ref{ssec:memb} and \ref{ssec:kin} (using a velocity window of \atevwindow{}) yields 
$\vsys=-328.0 \pm 2.3$ \kps{} and $\slos=8.1 \pm 1.8$ \kps. Member stars are 
clustered near the photometric center (Upper-right panel of Figure 
\ref{fig:and28kin}), and generally lie close to a fiducial 
isochrone for a dSph-like stellar population (Upper-right panel of 
Figure \ref{fig:and28kin}).  While there are a few stars with velocities
near the cold peak that appear to be outliers in the CMD, this is likely due 
to the relatively high photometric errors in this SDSS field.

The right panels of Figure \ref{fig:and28kin} reveal two stars categorized as members that are at large distances from the 
photometric center of \ate{}.  Based on the analysis described in \S 
\ref{ssec:memb} (assuming the surface brightness profile from  \citealt{Slater11And28}), there should be no \ate{} RGB stars at projected distances 
farther than these two stars  ($\sim 1100$ pc).  These two stars are also velocity outliers.  
Thus, either \ate{}'s surface brightness profile is intrinsically very different from other dSphs, 
the stars were tidally stripped, or these two stars are not associated with
\ate{}. Both stars' spectra have an absorption equivalent width of $\sim0.7$ \AA{} for Na I $
\lambda$8190, a surface-gravity sensitive feature that is only detected in dwarf stars.  This equivalent width marginally suggests they are 
dwarf stars, and hence unassociated foreground\footnote{For one of these stars 
the measurement may be contaminated by a sky line.}. While the foreground star estimate from \S \ref{ssec:memb} suggests that two MW 
foregrounds are unlikely, Poisson statistics imply a probability of $\sim 5\%$, so it is possible that both stars are contaminants.  

Hence, we also determine the kinematical parameters if these two stars are excluded, 
yielding $\vsys= -331.1 \pm 1.8$ \kps{} and $\slos=4.9 \pm 1.6$ \kps.  $\slos$ in this case
is significantly smaller than if these stars are included, but is stable within $1\sigma$ 
to the exclusion of any other two stars.   In the hypothesis that these stars are not members,
this velocity dispersion provides an  estimate of  \ate{}'s mass
(see \S \ref{ssec:mass} for details).  We combine this mass with the 3D distance from M31 ($367$ kpc) and a 
mass for M31 of $2 \times 10^{12} \Msun$ (intentionally in the high range of M31 masses) to estimate the Jacobi 
tidal radius of \ate{} \citep{bandt}.  

We find a  tidal radius for \ate{} of $\sim 3.4$ kpc, an order of magnitude larger 
than the current stellar extent ($R_e = 0.21$ kpc).  This strongly supports the hypothesis that the two outlier stars 
were \emph{not} recently  stripped.  Further, the $\vsys$ measured here for \ate{} is very close to the M31 
$\vsys$.  This may imply that it is near apocenter in its orbit, and thus is moving slowly  \citep[e.g.,][]{bk12leoi}.  
Hence, stars stripped at pericenter should be far from \ate{}.  Taken together, this  suggests \ate{} has not recently 
been tidally stripped, and the two anomalous stars are contaminants.   For our fiducial kinematic measurements of 
\ate{}  (listed in Table \ref{tab:props}), we exclude these two stars as likely contaminants.

In either case, as described below in \S \ref{ssec:mass}, this dispersion implies a mass-to-light ratio larger than any plausible stellar population.  Hence, we confirm that \ate{} is most likely a self-bound galaxy with a massive halo.   Its status as a satellite of M31 is examined in detail in \S \ref{ssec:satorfield}.

\subsection{\atn{}}

We present our spectroscopic measurements of stars in the \atn{} field along with the corresponding SDSS photometry in Figure 
\ref{fig:and29kin}.  As in \S \ref{ssec:ate}, a clear cold peak is present, but for \atn{}, the peak is closer to the MW peak at $\vh \sim -200$ \kps.  While \atn{} is closer to M31 than \ate{} ($\sim 190$ kpc projected), it is more offset from M31's $\vsys$, and is farther from the Galactic plane than 
\ate{} ($b \sim  -31 \deg$).  Thus both the MW and M31 contamination for the \atn{} field is  less than  \ate{}.
Indeed, applying our membership methodology to the cold peak (using a velocity window of \atnvwindow{}) provides the histogram shown in the lower-right panel of 
Figure \ref{fig:and29kin}.  This is consistent with a Gaussian distribution and has no obvious velocity outliers at large distances.
We measure its kinematical parameters as described in \S \ref{ssec:kin}, determine $
\vsys= -194.4 \pm 1.5$ \kps{} and $\slos=5.7 \pm 1.2$ \kps, and tabulate those results in Table \ref{tab:props}.  
As for \ate{}, these kinematical parameters are 
similar to those of other M31 and MW dSphs.  We conclude that \ate{}
is also a self-bound dSph satellite of M31.

\subsection{M31 Satellites or Local Group Field?}
\label{ssec:satorfield}

Both \ate{} \& \atn{} are at relatively large projected distances from M31 of $380$ and $190$ kpc, respectively.  By contrast, the majority of M31's 
dSphs lie within $150$ kpc projected. This large difference is primarily a selection effect caused by the angular extent of the PAndAS survey.  
Nevertheless, the large projected separation begs the question of whether or not \ate{} \& \atn{} are truly  satellites of M31 rather 
than ``free floating'' galaxies  of the LG (e.g., Tucana and Cetus).  While the typical usage of the word ``satellite'' for these galaxies is sometimes
ambiguous, here we define it to mean an object that is instantaneously bound to its host at the current epoch.

\begin{figure}[htbp!]
\epsscale{1.2}
\plotone{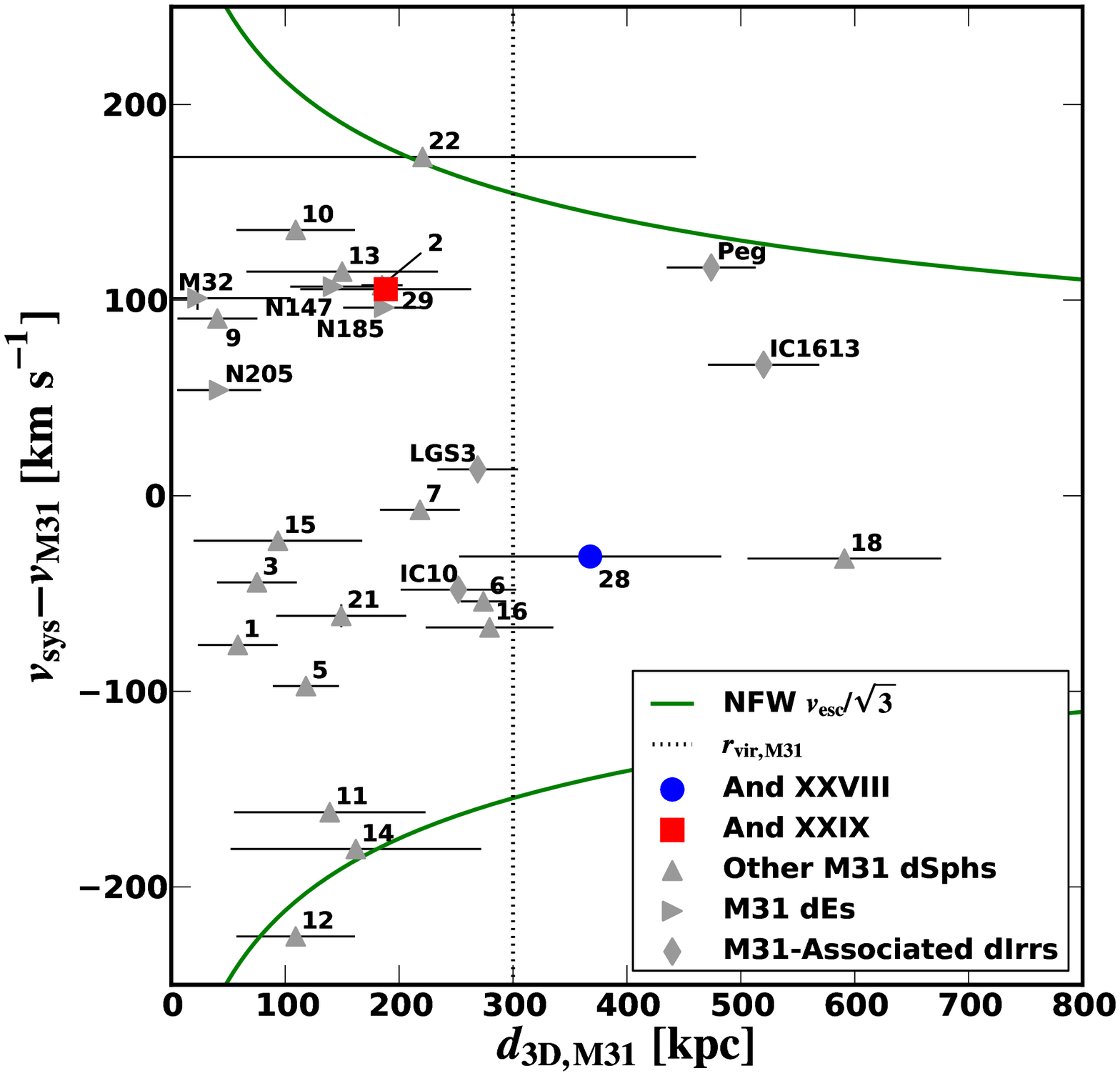}
\caption{Distance vs. line-of-sight $\vsys$ for M31-associated dwarf galaxies.  The (blue) circle and (red) square are from this work, \ate{} \& \atn, respectively.  Up-pointing triangles (gray) are dSphs from the samples of \citet{collins11and6}, \citet{paper1} and \citet{ho12}.  Diamonds and right-pointing triangles (gray) are dIrrs and dEs from \citet{mcc12LGcat}.  The distance axis is 3D distance from M31, computed by combining the projected separation of the dSphs from M31 with  line-of-sight TRGB distances.  The (black) vertical dotted line is the virial radius of M31 from the \citet{klypin02} model.  The (green) curves are the one-dimensional escape velocities from an NFW dark matter halo matching the model of \citet{klypin02}.   To make them one-dimensional, we divide the  escape velocities  by $\sqrt{3}$, so they represent the $\vsys$ above which a satellite would be unbound if its (unmeasured) tangential velocity components were equal to  $\vsys$.  \ate{} \& \atn{} lie well within the bound region. }
\label{fig:distvel}
\end{figure}

Figure \ref{fig:distvel} plots $\vsys$, the line-of-sight relative to M31, 
 against the 3D distance from M31 for a selection of dSphs.  
Also shown (as green curves) are the one-dimensional escape velocity 
(i.e., $v_{esc} / \sqrt{3}$) for the M31 halo model of \citet{klypin02} ($\mvir=1.6 \times 
10^{12} \Msun$). A satellite lying above this curve is thus unbound if its two 
tangential velocity components are equal to its observed line-of-sight $\vsys$.  It
is clear from inspection of this plot that both \ate{} (blue circle) \& \atn{} (red square) 
lie well within these 
curves, indicating that their tangential velocity components must exceed the 
line-of-sight velocity by a large margin to not be bound to M31.  This holds even
for \ate{}, despite the fact that it lies beyond M31's virial radius (vertical dashed line in Figure \ref{fig:distvel}.  

Also of note in Figure \ref{fig:distvel} is the satellite status of the most distant dSph, And XVIII. Its $\vsys$ 
is only  $\sim 30$ \kps{} from M31, despite being nearly as far behind M31 as M31 is from the MW.  While this strongly suggests it
is formally bound to M31, the free-fall time for And XVIII to reach M31 is greater than the time in which M31 is likely to merge with
the MW \citep{vdm12mwm31}. Hence, systems as distant from M31 (or the MW) as And XVIII might be better considered satellites
of the future merged M31/MW galaxy.

While the $\vsys$ measurements for these galaxies are suggestive, they cannot definitively determine if a satellite is bound 
due to the unmeasured tangential velocity.  However, the evidence that dark 
matter halos are present in both \ate{} \& \atn{} (\S \ref{ssec:mass}) provides strong constraints.
\citet{bk12leoi} shows that in $\Lambda$CDM simulations of either MW or 
M31-like halos, halos at distances from the host like those of \ate{} or \atn{} are nearly always 
($>99.9$ \%) bound.  In a $\Lambda$CDM context, it is thus extremely likely 
that \ate{} \& \atn{} are M31 satellites.

\subsection{Contamination Simulations}
\label{ssec:contamination}
The method described in \S \ref{ssec:kin} for estimating kinematical 
parameters assumes a pure sample of dSph member stars and does 
not formally account for the possibility of contamination due to MW 
foregrounds and M31 halo stars (discussed in \S \ref{ssec:memb}). 
To determine if such contamination might affect our kinematical results,
we simulate the effect of contamination on our parameter estimates by
creating mock velocity datasets designed to mimic \ate{} \& \atn{}. 

For each dSph we create 10000 mock datasets composed of a Gaussian
distribution with kinematic parameters corresponding to our results for the
dSph.  We add to this two additional uniform distributions with normalization set
to match the number of expected stars in the MW foreground and M31
halo as estimated in \S \ref{ssec:memb}.  We account for fractional numbers
of contaminants by only including an additional contaminant star for a 
corresponding fraction of the mock datasets. We then measure the
kinematical parameters of these mock velocity distributions, and compare them
to the input parameters used to generate the mock datasets. 

We generate mock datasets based on both the \ate{} \& \atn{} fields following the prescription described above.
For both sets of mocks, the estimated $\slos$ and $\vsys$ were very close to the true value used to
initialize the mock datasets. More specifically, the variance in the difference between the true and 
estimated values due to the contamination is $\sim 5$x lower than the variance in each measurement due to small number statistics.   
Thus, contamination at the level we estimate here cannot have a statistically significant impact on our results.

\begin{figure}[thbp!]
\epsscale{1.2}
\plotone{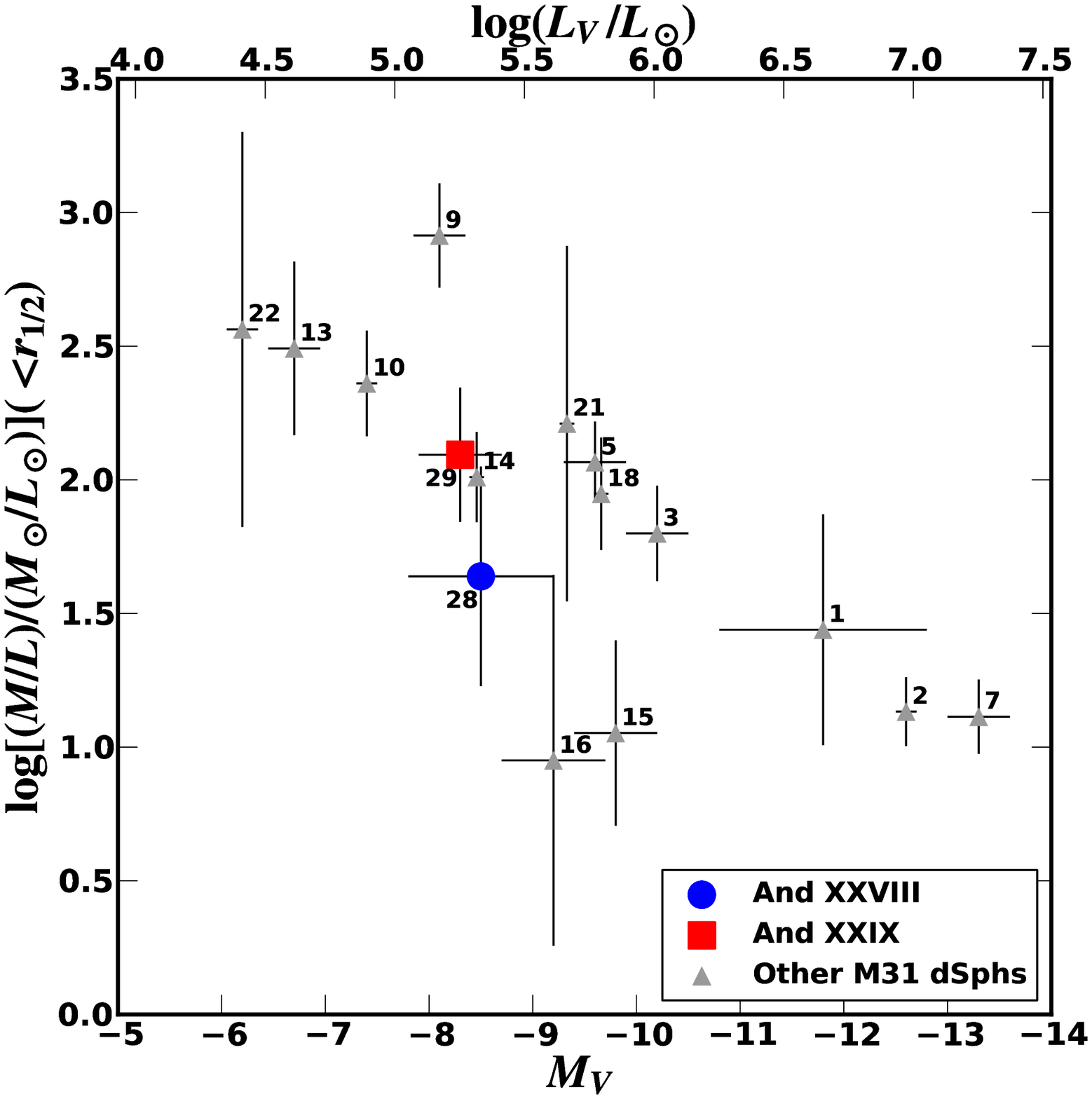}
\caption{Log of Mass-to-light ratio vs. Luminosity for M31 dSphs. \ate{} \& \atn{} are represented as a (blue) circle and (red) square, while other M31 dSphs from \citet{paper1} and \citet{ho12} are (gray) triangles.  Masses are determined using the \citet{wolf10} mass estimator (using half-light radii from  \citealt{brasseur11}), which is valid when interpreted as the mass within the deprojected (3D) half-light radius. Both \ate{} \& \atn{} lie in the range of other M31 dSphs and have mass-to-light ratios  greater than plausible  from their stellar populations alone, implying they are dark matter-dominated.}
\label{fig:mtol}
\epsscale{1}
\end{figure}

\subsection{Mass Estimates}
\label{ssec:mass}

With reliable kinematical parameters in hand, we are now in a position to estimate 
the mass of these galaxies from their internal velocities. We search for rotation by 
dividing the galaxies in half and search for mean velocities of opposite sign on each side, over a range of position angles. 
We find that the maximal rotation signal for either galaxy is negligible relative to the dispersion, $<1$ \kps.  
Hence we model the galaxies as purely pressure-supported systems.  We estimate the mass of these 
galaxies from their velocity dispersions ($M_{1/2}$) 
following Equation 2 of \citet{wolf10}, and tabulate these for \ate{} \& \atn{} in Table \ref{tab:props}.  
The mass obtained with this  formula is not strongly 
degenerate with the anisotropy, but in this interpretation, it is only valid as a mass 
measurement within the deprojected (3D) half-light radius\footnote{For surface brightness profiles appropriate for dSphs, however, the 
deprojected (3D) half-light radius is well  estimated as $4 R_{\rm eff} / 3$ where $R_{\rm eff}$ is 
the projected half-light radius.}. Applying this to our kinematical parameter estimates yields 
$M_{1/2}(<\rhalf)=4.7 \pm 3.2 \times 10^6 \Msun$ for \ate{}, and $M_{1/2}(<\rhalf)=1.1 \pm 0.49 \times 10^7 \Msun$ for \atn{}.

A simple mass-to-light ratio estimate may be 
obtained by  dividing this $M_{1/2}$  by the luminosity within $\rhalf$  --
by definition this is $L/2$.   For \ate{}, we obtain $M/L_V(<\rm r_{1/2})= 44 \pm 41 \; \Msun/\Lsun$, and for \atn{}, we find $M/L_V(< \rhalf)= 123 \pm 72 \; \Msun/\Lsun$.
  Like other M31 dSphs of similar luminosity (see Figure \ref{fig:mtol}), \ate{} \& \atn{} have mass-to-light ratios  above 
those that are possible for purely stellar systems.  This reveals the presence of a dark matter halo, even within the half-light radius, where 
baryons dominate for brighter galaxies \citep[e.g.,][]{stri08,stri08commonmass,walker09, tollerud11a}.  For \ate{}, the uncertainty admits  a stellar-like $M/L$ at the 
$1\sigma$ level, primarily due to uncertainty in the galaxy's luminosity, but the best estimate is well above any plausible stellar value. 
Alternatively, \citet{mcgaugh13andmond}, 
under a MOdified Newtonian Dynamics (MOND) hypothesis, have recently predicted velocity dispersions for \ate{} \& \atn{} of $4.3^{+0.8}_{-0.7}$ $4.1^{+0.8}_{-0.7}$ 
\kps{} (error bars are from assuming mass-to-light ratios from 1 to 4 $\Msun/\Lsun$). 
These predictions are marginally consistent with our measurements, but whether this is numerical coincidence given the uncertainties
  or evidence for MOND is beyond the scope of this paper \citep[][]{kaplinghat02, McGaugh11arxiv, foremanscott12, gnedin12}.

In Figure \ref{fig:mtol}, we show the mass-to-light ratios of \ate{} \& \atn{} in the context of other M31 dSphs.  
This plot demonstrates that their internal dynamics are fairly typical of M31's  dSphs, despite their distance from their host.  
This implies that the processes shaping these dSphs' kinematic structure either operate quickly upon infall, are in place before the 
satellites interacted with M31, or both \ate{} \& \atn{} have been in the M31 system for some time and have already been 
 influenced by M31's environment \citep[e.g.,][]{wetzel12}. These trends are also similar to MW dSphs, suggesting these processes are not 
 unique to the M31 system \citep{paper1}.

\section{Conclusions}
\label{sec:conc}

The spectroscopic observations of the \ate{} \& \atn{} fields described here support the following conclusions:

\begin{enumerate}

\item We have spectroscopically confirmed that \ate{} \& \atn{} are likely self-bound dwarf galaxies, with velocity dispersions of $4.9 \pm 1.6$ and $5.7 \pm 1.2$ \kps, respectively.  The implied large mass-to-light ratios are consistent with dark matter-dominated dynamics.

\item While they are in the outskirts of the M31 system, the systemic velocities of \ate{} \& \atn{} imply they are both bound to M31.

\item The internal kinematics and implied masses of \ate{} \& \atn{} are like those of other dSphs of similar luminosities, despite the large distances from their host.  

\end{enumerate}

While these results clearly demonstrate that \ate{} \& \atn{} are self-bound dwarf satellites of M31, their nature and evolutionary history are still 
open questions.  \atn{} seems to have only older ($\gtrsim$ a few Gyr) stars \citep{Bell11And29}, but \ate{}'s stellar population is less constrained by 
the SDSS photometry \citep{Slater11And28}.  The paucity of stars near \ate{} with  $g-r<0$ or $r-i<0$ implies it cannot have as much recent star 
formation as Leo T, but low levels of intermediate-age stars cannot be completely ruled out without deeper photometry. Further, without HI data, it 
cannot be definitively determined if these galaxies are truly passive dSphs, or a ``transition''-type dSph/dIrr objects in a low star-formation episode.

The lack of recent star formation in \ate{} is surprising, as it stands in contrast to galaxies at similar distances from their hosts like Leo T and 
the Phoenix Dwarf.  Yet unlike Tucana and Cetus, \ate{}  is unambiguously associated with M31.  Thus, \ate{}  stands as an important data point 
for understanding the evolutionary history of dSphs satellites and their connection to field galaxies.

\acknowledgements{The authors acknowledge Sarah Benjamin, Ana Bonaca , Nitya Kallivayalil, and Nhung Ho for valuable feedback on this manuscript. We also thank the referee (Alan McConnachie) for helpful feedback that improved this work.

Support for this work was provided by NASA through Hubble Fellowship grant \#51316.01 awarded by the Space 
Telescope Science Institute, which is operated by the Association of Universities for Research in Astronomy, Inc., for NASA, under contract NAS 
5-26555.

The authors wish to recognize and acknowledge the very significant cultural role and reverence that the summit of Mauna Kea has always had 
within the indigenous Hawaiian community.  We are most fortunate to have the opportunity to conduct observations from this mountain.

Funding for SDSS-III has been provided by the Alfred P. Sloan Foundation, the Participating Institutions, the National Science Foundation, and 
the U.S. Department of Energy Office of Science. The SDSS-III web site is http://www.sdss3.org/.

SDSS-III is managed by the Astrophysical Research Consortium for the Participating Institutions of the SDSS-III Collaboration including the 
University of Arizona, the Brazilian Participation Group, Brookhaven National Laboratory, University of Cambridge, Carnegie Mellon University, 
University of Florida, the French Participation Group, the German Participation Group, Harvard University, the Instituto de Astrofisica de 
Canarias, the Michigan State/Notre Dame/JINA Participation Group, Johns Hopkins University, Lawrence Berkeley National Laboratory, Max 
Planck Institute for Astrophysics, Max Planck Institute for Extraterrestrial Physics, New Mexico State University, New York University, Ohio State 
University, Pennsylvania State University, University of Portsmouth, Princeton University, the Spanish Participation Group, University of Tokyo, 
University of Utah, Vanderbilt University, University of Virginia, University of Washington, and Yale University.}

This research made use of Astropy (\url{http://www.astropy.org}), a community-developed core Python package for Astronomy.

{\it Facilities:}  \facility{Keck:II (DEIMOS)}, \facility{APO (SDSS)}

\object{And XXVIII}, \object{And XXIX}, \object{M31}

\bibliography{and2829}{}
\bibliographystyle{apj}

\end{document}